\documentclass{article}

\usepackage{microtype}
\usepackage{graphicx}
\usepackage{subcaption}
\usepackage{booktabs} 
\usepackage{svg}

\usepackage{xurl}
\usepackage{hyperref}

\usepackage[preprint]{icml2026}

\usepackage{amsmath}
\usepackage{amssymb}
\usepackage{mathtools}
\usepackage{amsthm}
\usepackage{caption}

\usepackage[capitalize,noabbrev]{cleveref}

\theoremstyle{plain}

\theoremstyle{definition}

\theoremstyle{remark}

\usepackage[textsize=tiny]{todonotes}

\icmltitlerunning{Does Persona Make LLMs K-pop Fans?}

\begin{document}

\twocolumn[
  \icmltitle{Does Persona Make LLMs K-pop Fans? \\
  A Pilot Study of LLM-Based Online Concert Audience Agents}



  \icmlsetsymbol{equal}{*}

  \begin{icmlauthorlist}
    \icmlauthor{Kirak Kim}{yyy}
    \icmlauthor{Hyojin Kim}{yyy}
    \icmlauthor{Yejin Son}{xxx}
    \icmlauthor{Sungyoung Kim}{yyy}
    \icmlauthor{Kyung Myun Lee}{yyy}
  \end{icmlauthorlist}

  \icmlaffiliation{yyy}{Graduate School of Culture Technology, KAIST, Daejeon, South Korea}
  \icmlaffiliation{xxx}{Department of Artificial Intelligence, Yonsei University, Seoul, South Korea}
  
  \icmlcorrespondingauthor{Kyung Myun Lee}{kmlee2@kaist.ac.kr}
  \icmlcorrespondingauthor{Sungyoung Kim}{sungyoung.kim@kaist.ac.kr}

  \icmlkeywords{LLM Agents, Online Concert, Audience Simulation, Cultural AI Evaluation, Human-AI Interaction}

  \vskip 0.3in

]



\newcommand{\blue}[1]{\textcolor{blue}{#1}}
\newcommand{\cmt}[1]{\textcolor{red}{#1}}

\printAffiliationsAndNotice{}  


\begin{abstract}
A concert is a collective experience, but recorded performance videos are typically watched alone, stripping away the shared audience presence that makes concerts feel eventful. We investigate whether persona-based LLM audience agents can recreate aspects of this collective experience by generating real-time fan chat alongside a K-pop performance video. We present a multi-agent system in which ten LLM agents react through live-chat messages, comparing a persona-conditioned audience (each agent assigned a distinct fan identity, bias, and chat style) with a no-persona baseline. In a within-subjects pilot with K-pop fans (N=11), persona conditioning substantially improved model-level chat quality and perceived naturalness, but did not translate into differences in social connectedness, engagement, or affective response. Interviews suggest that online K-pop concert chat may operate as collective monologue rather than interpersonal dialogue, and that meaningful participation depends on shared identification with the specific artist and fandom. Persona conditioning can make LLM audiences appear more natural, but culturally meaningful collective experience may require deeper alignment between persona, crowd behavior, fandom identity, and user expectations.
\end{abstract}

\section{Introduction}
 
A concert is a collective experience. What distinguishes a live performance from a recording is not merely acoustic or visual fidelity, but the shared presence of an audience whose synchronized responses (e.g., cheers, chants) help constitute the event itself~\cite{auslander2022liveness}. As concerts increasingly move online, livestreams and VR concert platforms have expanded access to performances for audiences who may not be able to attend in person. Yet this access often comes at the cost of collective experience. Outside synchronously scheduled livestreams, recorded performance videos are typically watched alone, stripping away the sense of shared audience presence that makes concerts feel eventful. Empirical work on synchronous online concerts has shown that text-chat interaction is itself a primary medium through which audiences' co-presence and collective affect are produced ~\cite{kim2025exploring}. In this sense, chat is not merely incidental to online concert experience but a constitutive part of how viewers feel together.

This paper explores whether persona-based LLM audiences can meaningfully enhance users’ experience of asynchronous online concert viewing. We develop an LLM-based audience system for K-pop performance videos, where multiple chat agents react moment-by-moment using the expressive conventions of online K-pop fan culture. Because K-pop fandom is highly individuated---fans differ in their bias members, engagement levels, chat styles, and relationships to the group---we investigate \textbf{persona} conditioning as a design strategy for producing a more diverse and believable AI audience.

We evaluated this assumption through a within-subjects pilot study with K-pop fans (N=11), comparing a persona-conditioned audience of distinct fan agents with an unconditioned baseline. Our findings reveal an unexpected gap between model-level output quality and user-level experience. Persona conditioning substantially differentiated agent behavior and reduced homogeneous, repetitive chat patterns. However, these differences did not translate into stronger social connectedness, engagement, or affective response; only perceived naturalness showed a noticeable advantage. Interviews suggest that this gap arises from cultural properties of K-pop fan chat. These findings complicate the assumption that improving the perceived naturalness of persona-based AI agents necessarily translates into more meaningful collective experiences, and point toward fandom-specific identity alignment as a persona design challenge as well as the need for culturally situated, interpretive evaluation of cultural LLM agents~\cite{kommers2026computational}.

\section{Related Works}
\subsection{Online Concert Experience}
Prior research has highlighted the importance of social experiences in online concerts and live-streamed performances~\cite{ppali2025virtual}. In particular, social presence and connectedness have been identified as key factors shaping users’ engagement and overall experience in live music performance \cite{koefler2026let}. These social aspects are especially prominent in K-pop fandom culture, where audiences actively participate in collective practices such as fan chants, sing-alongs, and real-time interactions on social media platforms~\cite{upham2024audience}. Accordingly, recent studies have explored ways to support collective and socially engaging experiences in online concert platforms across various contexts, including K-pop performances~\cite{kim2025bring}. Similar efforts have also extended to asynchronous video platforms by replaying reactions from previous audiences. For example, \textit{danmaku} displays comments from previous viewers synchronized with the video timeline directly on the screen~\cite{chen2017watching,wu2019danmaku}. Another system transformed audience chat and reaction data from existing videos into the behaviors of virtual avatars in VR concert environments~\cite{lee2025concert}. 

These previous approaches have primarily relied on replaying past users' data. However, little attention has been paid to generating virtual audience responses using AI. Systems based on existing user data are inherently limited, as they require a sufficient amount of pre-collected audience reactions to function effectively. LLM-based virtual audience could address this limitation by generating new reactions dynamically, without relying solely on previously accumulated data. Therefore, in this work, we explore whether LLM-generated K-pop fans can participate in online concerts and foster social experiences for users.

\subsection{Persona-Based LLM Agents}

Recent research has explored persona-based LLM agents across a range of applications and interaction settings. A line of research has used persona-based LLM agents as simulated participants, aiming to approximate population-level feedback across diverse demographic, behavioral, or preference profiles~\cite{argyle2023out, manning2024automated}. In this approach, LLMs are conditioned on personas that encode various human attributes, such as demographic information, personality traits, and preferences, in order to simulate responses from a society or population. While this approach has shown some promise across multiple domains, recent work suggests that richer personas or more detailed synthetic profiles do not necessarily lead to more realistic population simulation; instead, they may amplify demographic biases or stereotypical assumptions~\cite{li2026llm}. This motivates careful, context-grounded persona design grounded in detailed analysis of the target domain.

Persona and role-playing have also been studied in the context of character simulation and interactive storytelling in virtual worlds~\cite{park2023generative, shao2023character, wang2025characterbox}. In sandbox environments or text-based virtual worlds, LLM agents are assigned specific characters and roles, and researchers examine what kinds of social behaviors emerge through their interactions. These studies focus on how persona-conditioned agents continuously interact with their environment and with other agents, forming social trajectories over time. Inspired by these lines of work, we assign each LLM agent a distinct fan persona and treat K-pop fan chat not as a set of isolated utterances, but as an emergent group-level response produced by multiple persona-conditioned agents.
 
\section{LLM-Based Audience Design}
\label{sec:method}

\subsection{Overview}

We propose a multi-agent LLM-based audience simulation framework grounded in empirical observations of K-pop live-chat behavior and performance segment labels. The system combines prompt design informed by observed fan-chat practices with timestamp-aligned segment labels, allowing agent utterances to remain temporally synchronized with the stage and stylistically anchored in real fan-chat conventions. The system is deployed as a real-time web application in which a participant watches the performance video alongside ten LLM-driven audience agents and can freely participate in the chat with them.

For this pilot study, we used K-pop girl group LOONA's \textit{Butterfly} stage (aired 2019.02.21, duration $\approx$4~min) as the target video for the audience simulation\footnote{\url{https://youtu.be/_wNEPao6lXU?si=bUEFtNnMFVj6mje5}}. The video was manually annotated by a human labeler with timestamp-aligned segment labels describing the musical structure, active vocalists, and key choreographic events at each moment. We chose this video in part because LOONA's fandom, \textit{Orbits}, is internationally distributed with a strong English-language online presence, and an active culture of memes and live-chat engagement~\cite{russell2019stanloona}. This makes chat-style fan reactions both plausible and relatively well documented, and such material may also be reflected in the web-scale data on which LLMs are trained.

\subsection{Live-Chat Analysis for System Prompt Design}
\label{sec:empirical}

To ground the agents' chat behavior in observed K-pop fan practices, we analyzed YouTube live-chat logs collected from an online K-pop performance video\footnote{\url{https://youtu.be/w89rNW8H_1k?si=FkSdmt83CgsWCV9-}}. Adapting prior coding schemes for music-audience comments~\cite{lamont2017music, fraser2021music}, we identified four primary categories: 1) \textbf{interaction} (communication with the artist or other audience members, e.g., shout-outs, compliments), 2) \textbf{collective reaction} (audience behaviors common in offline concerts, e.g., applause, cheering), 3) \textbf{personal thought} (self-directed comments, e.g., preferences, nostalgia), and 4) \textbf{emotion} (affective responses, often via emojis or short bursts). In addition to these four, we treated two recurring patterns as design-relevant observations, \textbf{repetitive bias-name chants} and \textbf{first-person questions}.




\subsection{Generation Framework}
\label{sec:generation}

Our framework models audience behavior as a per-agent, context-conditioned generation process. At each generation step $t$ (every 8 seconds), the system retrieves a \textbf{segment label} $S_t = \{\text{section},\ \text{vocals},\ \text{description}\}$ where \textit{section} denotes the musical section at $t$ (e.g., \texttt{verse\_1}, \texttt{chorus\_2}, \texttt{bridge}), \textit{vocals} is the list of active vocalists, and \textit{description} is a prose narrative of the visual and choreographic event at that moment.


Alongside the segment label, each generation call includes a \textbf{phase guide} $G_t \in \{\text{preshow}, \text{performance}, \text{postshow}\}$, a short style-prompt with phase-appropriate example utterances (prompts in Appendix~\ref{app:phase-guides}), and a \textbf{moment cue} $M_{i,t}$. The moment cue highlights salient events in the current segment, including chorus sections, key choreographic moments, and the closing section.


Each agent additionally receives two conversational memory components: 1) a \textbf{personal history buffer} $\mathcal{H}_{i,t}$ containing its own recent messages (last 6), which discourages self-repetition, and 2) a \textbf{shared chat log} $\mathcal{C}_t$ containing recent messages from all agents and the human participant (last 10 entries), enabling inter-agent and agent--participant interaction.

For each agent $i$, these components are assembled into an agent-specific prompt context

\begin{equation}
X_{i,t} = \{ P_i,\ S_t,\ G_t,\ M_{i,t},\ \mathcal{H}_{i,t},\ \mathcal{C}_t \},
\end{equation}

where $P_i$ denotes the \textbf{persona specification} of agent $i$.
The LLM maps this context to a structured response
$f_\theta(X_{i,t}) = \{z_{i,t}, m_{i,t}\}$,
where $z_{i,t} \in \{0,1\}$ denotes the binary speaking decision and
$m_{i,t}$ is a candidate chat message.

To approximate natural chat timing, each message is released with 2--8 seconds of uniform-random jitter after inference completes. Generation is initiated 8 seconds ahead of the session clock so that messages are typically ready on time even under inference latency.

All ten agents are batched into a single call per generation step. Agent responses are generated using Qwen2.5-32B-Instruct-AWQ and sampling temperature differs by persona condition (Section~\ref{sec:personas}).

\subsection{System Architecture}

The backend is implemented as a FastAPI application with WebSocket support. The human participant joins via WebSocket, can send their own chat messages, and these messages are injected into the shared context $\mathcal{C}_t$ so that agents can respond in subsequent generation cycles.

\subsection{Persona Design}
\label{sec:personas}

We define ten audience agents along five dimensions: \textbf{age}, \textbf{gender}, \textbf{region}, \textbf{bias (favorite member)}, and \textbf{chat style}. The resulting personas span a range of engagement modes (from high-frequency bias-locked fans to first-time viewers without bias) grounded in the chat patterns identified in Section~\ref{sec:empirical}. Each agent's persona text is concatenated to a shared global prompt that establishes the task framing, output format constraints, and persona-consistency rules (both the global prompt and the full per-agent persona texts are provided in Appendix~\ref{app:prompts}).

To evaluate the contribution of persona design, we include a \textbf{no-persona} baseline in which all ten agents are stripped of individual identity, bias, and behavioral traits: each agent's persona text is set to an empty string and its nickname is reduced to the bare first name with no other identifying information. The shared global prompt, which establishes the LOONA-fan task framing, chat-format constraints, and a small set of phase-specific few-shot fan-chat utterances used to anchor chat-style output, is retained in both conditions. Crucially, few-shot K-pop fan-chat examples appear only in the shared global prompt and are not embedded in the per-agent persona texts. Both conditions receive identical few-shot anchoring material, preventing differences in response quality from being attributed to unequal access to example utterances rather than to per-agent persona conditioning. Sampling temperature was raised from 0.7 to 1.2 in the no-persona baseline to compensate for output collapse (Appendix~\ref{app:no-persona}).


\section{User Study}

\subsection{Participants}

Eleven Korean K-pop fans fluent in English (8 women, 3 men; $M_{\text{age}}=24.7$, $SD=1.34$) participated in the pilot user study. None identified as members of LOONA's fandom, \textit{Orbits}. The chat system operated in English and participants interacted in English throughout the session. 

\subsection{Design}

We employed a within-subjects design with two conditions: 1) a \textbf{persona-conditioned} LLM audience and 2) a \textbf{no-persona} baseline. Condition order was randomized across participants. The performance stimulus (LOONA's 2019 \textit{Butterfly} stage) was identical in both conditions. The experimental interface is shown in Appendix~\ref{app:screenshot}.

\subsection{Procedure}

Each session lasted approximately 30 minutes and proceeded in five phases:

\begin{enumerate}
    \item \textbf{Pre-experiment questionnaire}: assessing participants' K-pop fandom background (favorite groups, fandom duration, engagement style) and prior experience with online concert formats (livestreams, VR concerts).
    \item \textbf{Familiarization viewing}: watching the \textit{Butterfly} video once with no chat present, equalizing first-exposure novelty across the two experimental conditions.
    \item \textbf{Condition 1 \& 2}: viewing of the stage with the audience condition (persona and no-persona), followed by the post-condition questionnaire.
    \item \textbf{Semi-structured interview}: asking open-ended questions about overall impressions, motivation to participate, perceptions of artificiality, and whether distinct characters were perceived among the agents.
\end{enumerate}

\subsection{Measures}

After each condition, participants completed four measures:

\begin{itemize}
    \item \textbf{Inclusion of Other in the Self} (IOS) \cite{aron1992inclusion}: a single-item pictorial measure of perceived social connectedness, indexing felt bonding with the LLM audience as a proxy for collective experience.
    \item \textbf{User Engagement Scale---Short Form} (UES-SF) \cite{o2018practical}: a 12-item measure of user engagement during the experience, covering focused attention, perceived usability, aesthetic appeal, and reward.
    \item \textbf{Self-Assessment Manikin} (SAM) \cite{bradley1994measuring}: pictorial scales for affective valence, arousal, and dominance.
    \item \textbf{Perceived naturalness}: a 1–7 Likert item asking how natural the LLM audience felt as a concert audience.
\end{itemize}
Given the small sample size, we report descriptive statistics and effect sizes (Cohen's $d_z$ for within-subjects comparisons) alongside null-hypothesis tests.

\subsection{Interview Analysis}
We conducted thematic analysis using a bottom-up approach \cite{braun2006using}. One author initially reviewed the interview transcripts, identified meaningful excerpts, and generated initial codes inductively from the data. The authors then discussed and grouped related codes into four main themes reflecting participants’ experiences and perceptions of the system.

\section{Results: The Divergence of Output and Experience}
\label{sec:results}
\subsection{Multi-level Chat Evaluation Metrics}
We analyzed all 22 sessions (11 participants $\times$ 2 conditions) using the following metrics applied to the agent chat output. \emph{Output-level} metrics treat each session's chat as a single corpus: lexical diversity is measured by Distinct-2 (bi-grams divided by total) and by Self-BLEU at n-gram order 2 (each message scored as a hypothesis against all other messages as references and then averaged across messages; lower indicates more diverse output)~\cite{zhu2018texygen}; output collapse by the exact-repetition rate (the fraction of messages that verbatim duplicate an earlier message in the session); output volume by mean message length. \emph{Per-agent profile spread} measures how behaviorally differentiated the ten simulated agents are from one another within a single session: for each per-agent metric (exact-repetition rate, mean message length, emojis-per-message, all-caps rate, and message count), we compute the range (max $-$ min) across the ten agents. High spread indicates persona-distinct simulated participants; low spread indicates agents collapsing to a near-identical profile.

For model-level metrics computed on the agent chat, we report paired comparisons between conditions. Because all model-level metrics showed consistent directional effects across participants, we emphasize Cohen's paired $d_z$ to index effect magnitude. For the post-session questionnaire battery, we report Mann--Whitney $U$, two-sided $p$, and rank-biserial $r$.
\subsection{Model-Level Differentiation}
\paragraph{Persona conditioning produces dramatic model-level differentiation}
\label{sec:results-model}

\begin{figure*}[t]
    \centering
    \includegraphics[width=\textwidth]{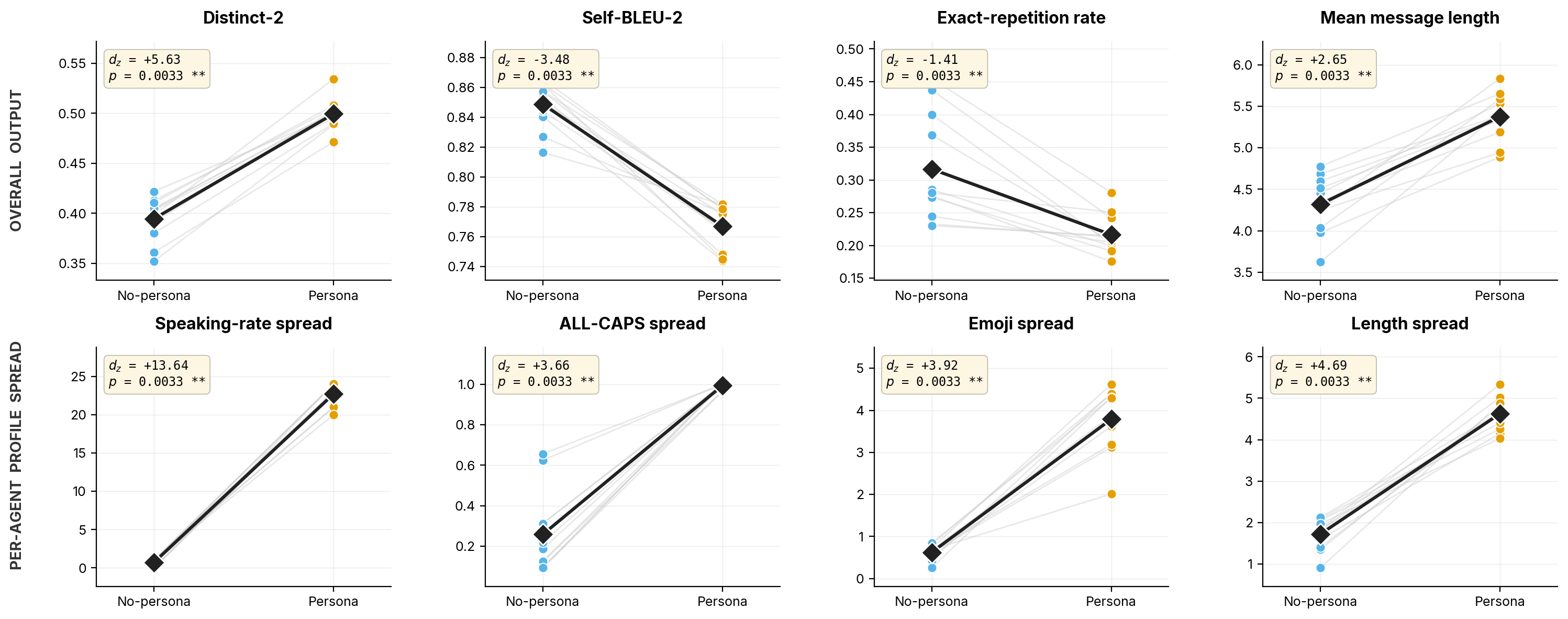}
    \caption{
        Model-level effects of persona conditioning. Each thin line represents one participant's paired sessions across conditions, while black diamonds indicate group means. Persona conditioning produces strong model-level differentiation.
    }
    \label{fig:model_effects}
\end{figure*}
 
\autoref{fig:model_effects} summarizes eight key metrics, organized into \emph{overall output} properties of the chat as a whole (top row) and \emph{per-agent profile spread}, the range of each behavioral metric across the ten simulated agents within one session (bottom row).
 
Persona conditioning increases overall output diversity (Distinct-2: $d_z = +5.63$; Self-BLEU-2: $d_z = -3.48$, where lower is more diverse), reduces verbatim repetition (exact-repetition rate: $d_z = -1.41$), and yields longer messages on average (mean message length: $d_z = +2.65$). The qualitative form of the no-persona collapse is illustrative. In one session, ten agents simultaneously produced the message ``\textit{orbits assemble}'' before the show, directly echoing an example from the system prompt; identical bursts recurred throughout the session, with the same agents independently re-converging on prompt examples several times. Persona conditioning eliminates this duplicate-cascade behavior.
 
The most distinctive finding lies in per-agent profile spread. Without persona, the ten agents converge to a near-identical behavioral profile across every dimension we measured: similar mean message length, similar use of all-caps, similar use of emojis, and an essentially deterministic speaking rate of approximately 32 messages per session (the maximum possible, equal to one message per generation turn). Persona conditioning differentiates the ten agents simultaneously along all of these dimensions. The most extreme effect is in \emph{speaking-rate spread} ($d_z = +13.64$): the persona condition produces a median range of 23 messages between the loudest and quietest agent, against 1 message in the no-persona condition. Length spread ($d_z = +4.69$), emoji spread ($d_z = +3.92$), and ALL-CAPS spread ($d_z = +3.66$) show similarly large effects. We read this cluster of metrics as direct evidence that persona conditioning produces not merely \emph{more diverse} outputs but \emph{behaviorally distinct} simulated participants.

\begin{table}[t]
\caption{Per-agent speaking rates by condition (mean $\pm$ SD across $N=11$ paired sessions. Under No-persona, all ten agents speak at the ceiling rate of $\sim$32 messages (one per generation turn); under Persona, speaking rate becomes a design-aligned signature ranging from $9.2$ to $31.8$. Full persona prompts in Appendix~\ref{app:personas}.}
\label{tab:speaking_rate}
\centering
\small
\setlength{\tabcolsep}{6pt}
\begin{tabular}{lcc}
\toprule
Agent persona (bias)     & Persona        & No-persona     \\
\midrule
Gowon lurker (Gowon)       & $9.2 \pm 1.5$  & $31.9 \pm 0.3$ \\
Yves-only stan (Yves)      & $11.7 \pm 1.0$ & $31.9 \pm 0.3$ \\
Heejin chant (Heejin)      & $20.5 \pm 2.0$ & $31.7 \pm 0.5$ \\
Kim Lip casual (Kim Lip)   & $21.6 \pm 1.9$ & $31.8 \pm 0.4$ \\
Dance nerd (---)           & $26.7 \pm 0.9$ & $31.6 \pm 0.5$ \\
Choerry all-caps (Choerry) & $28.6 \pm 1.2$ & $31.9 \pm 0.3$ \\
Haseul sentimental (Haseul)& $30.6 \pm 0.7$ & $32.0 \pm 0.0$ \\
Newbie (---)               & $30.8 \pm 0.6$ & $31.6 \pm 0.5$ \\
Lore explainer (---)       & $31.7 \pm 0.9$ & $31.9 \pm 0.3$ \\
JinSoul emoji (JinSoul)    & $31.8 \pm 0.6$ & $32.0 \pm 0.0$ \\
\bottomrule
\end{tabular}
\end{table}
 
\paragraph{Persona signatures alignment with design intent}
\label{sec:results-design}
 
\autoref{tab:speaking_rate} disaggregates the speaking-rate effect by agent. The two agents with explicit identity-grounded \emph{lurker} prompts (the Gowon-only fan, who ``only breaks for bias'', and the Yves-only fan, who is ``quiet when she is off-screen'') produce the most dramatic suppression (mean $= 9.2$ and $11.7$ messages, respectively, versus ${\sim}32$ in no-persona). Selective-reactor agents (Heejin-chant fan, Kim~Lip casual viewer) cluster in the middle (${\sim}21$ messages). Agents whose prompts permit speaking on every cue (lore explainer, JinSoul-emoji ``hypes everyone'') match the no-persona ceiling.
 
However, this alignment is not consistent across cue types. Identity-grounded cues (``Gowon lurker, only breaks for bias'') suppress speaking strongly; behaviour-instructional cues (``type rarely'', ``does not spam'') barely suppress speaking (the newbie agent, prompted to ``type rarely'', produced $30.8$ of $32$ possible messages on average.)
 
\paragraph{Subjective experience does not track model-level differentiation}
\label{sec:results-experience}
 
Across the post-session battery (IOS, UES-SF subscales, SAM, naturalness), only naturalness differed between conditions ($M_{\text{persona}} = 5.45$, $M_{\text{no-persona}} = 4.00$; $U = 91.0$, $p = .045$, $r = .43$). All other measures yielded $|r| < 0.15$ and non-significant $p$ (full subjective evaluation results are in Appendix~\ref{app:questionnaire}). Participants therefore detected the naturalness difference between conditions, yet this detection did not translate into differentiated experience along any other dimension we measured.

\subsection{User-Level Experience}
To understand why model-level differentiation registered subjectively only as naturalness, we analyzed post-session interviews. Among the four themes identified through thematic analysis, we report the three most relevant to explaining the quantitative findings (full themes in Appendix~\ref{app:interview_quotes}).
\paragraph{Collective audience dynamics contributed more to perceived realism than individual personas.} The majority of participants (N=8) reported that they could not clearly perceive distinct individual personas within the audience. One participant explained, ``I don’t think I perceived the audience members as separate individuals. I mainly focused on the overall atmosphere and trends of the chat.'' (P05) Participants also frequently described realism in terms of collective chat dynamics, such as repetitive comments, synchronized reactions, and spam-like audience behaviors. This may reflect the nature of real-world live-chat environments, where interactions are often dominated by parallel emotional expressions, artist-focused reactions, and crowd-like behaviors rather than direct interpersonal communication between audience members.

Therefore, while persona-based audience generation improved perceived naturalness by making individual members appear more human-like, it did not substantially enhance higher-level dimensions such as social connectedness or emotional engagement. Bridging this gap will likely require deeper study of fandom and online chat culture, and analysis of how individual fan traits aggregate into group-level interaction patterns. Persona design grounded in such understanding may meaningfully shift collective audience dynamics.


\paragraph{Shared cultural context influenced engagement.} Although participants were familiar with K-pop performances and live-chat culture in general, several reported difficulty engaging with the AI audience because they were unfamiliar with the specific artist and fandom represented in the chat. One participant noted, ``Since I was not a fan of the artist, I did not really know what to say. If it had been a video of an artist I personally liked, I think I would have felt more motivated to participate in the chat. Here, it was difficult to respond because I could not match the members’ names with their faces'' (P08). Participants also mentioned language barriers and unfamiliar fandom-specific expressions as factors limiting participation.

Nevertheless, the \textbf{chat interactions still fostered a sense of co-presence and collective engagement.} More than half of the participants (N=6) reported feeling as if they were watching the performance together with others, and most participants participated in the ongoing chat interaction (N=10). These findings suggest that our system was able to evoke a lightweight form of social presence. Interestingly, several participants reported experiencing audience presence in both conditions, indicating that even the no-persona condition was sufficient to create a basic sense of collective viewing. This aligns with the quantitative results showing no significant difference in social connectedness scores between conditions. While participants appeared to experience a certain level of co-presence, enhancing individual agent realism alone may not have been sufficient to substantially increase interpersonal connectedness.

\section{Conclusion}

This pilot study examined whether persona-conditioned LLM agents can make asynchronous online concert viewing feel more like watching with a live K-pop audience. Our results show that persona conditioning substantially improved the generated audience at the model-output level. This improvement was also perceptible to participants, as the persona condition was rated as significantly more natural. However, this increase in perceived naturalness did not translate into significant differences in social connectedness, engagement, or affective response. Interview findings suggest that this gap may stem from two cultural properties of online K-pop concert chat: first, it often functions less as interpersonal dialogue than as collective monologue, where participants attend to the atmosphere of the crowd rather than to distinct individuals; second, meaningful participation depends on shared cultural context and in-group identification with the specific artist and fandom. These findings suggest that current persona design is a useful but insufficient strategy for LLM-based audiences.

In future work, we plan to evaluate the system with fans of the performing artist to examine whether stronger fandom alignment allows perceived naturalness to translate into social and emotional experience. We also plan to investigate how disclosure of AI mediation shapes participation, since some participants reported that knowing the audience was AI-generated was itself a novel experience and reduced the social pressure of chatting. Overall, our findings indicate that culturally meaningful collective experience may require deeper alignment between persona, crowd behavior, fandom identity, and users’ expectations of AI audiences.

\section*{Impact Statement}

LLM-based audience agents may help support social presence in asynchronous online concert viewing, but they also raise concerns around transparency and cultural representation of fandom culture. We argue that such systems should clearly disclose AI mediation and be designed with care for the norms, identities, and expectations of the communities they simulate.


\bibliography{mybib}
\bibliographystyle{icml2026}

\newpage
\appendix
\onecolumn

\clearpage
\section{System Interface Screenshot}
\label{app:screenshot}

\begin{center}
    \includegraphics[width=\textwidth]{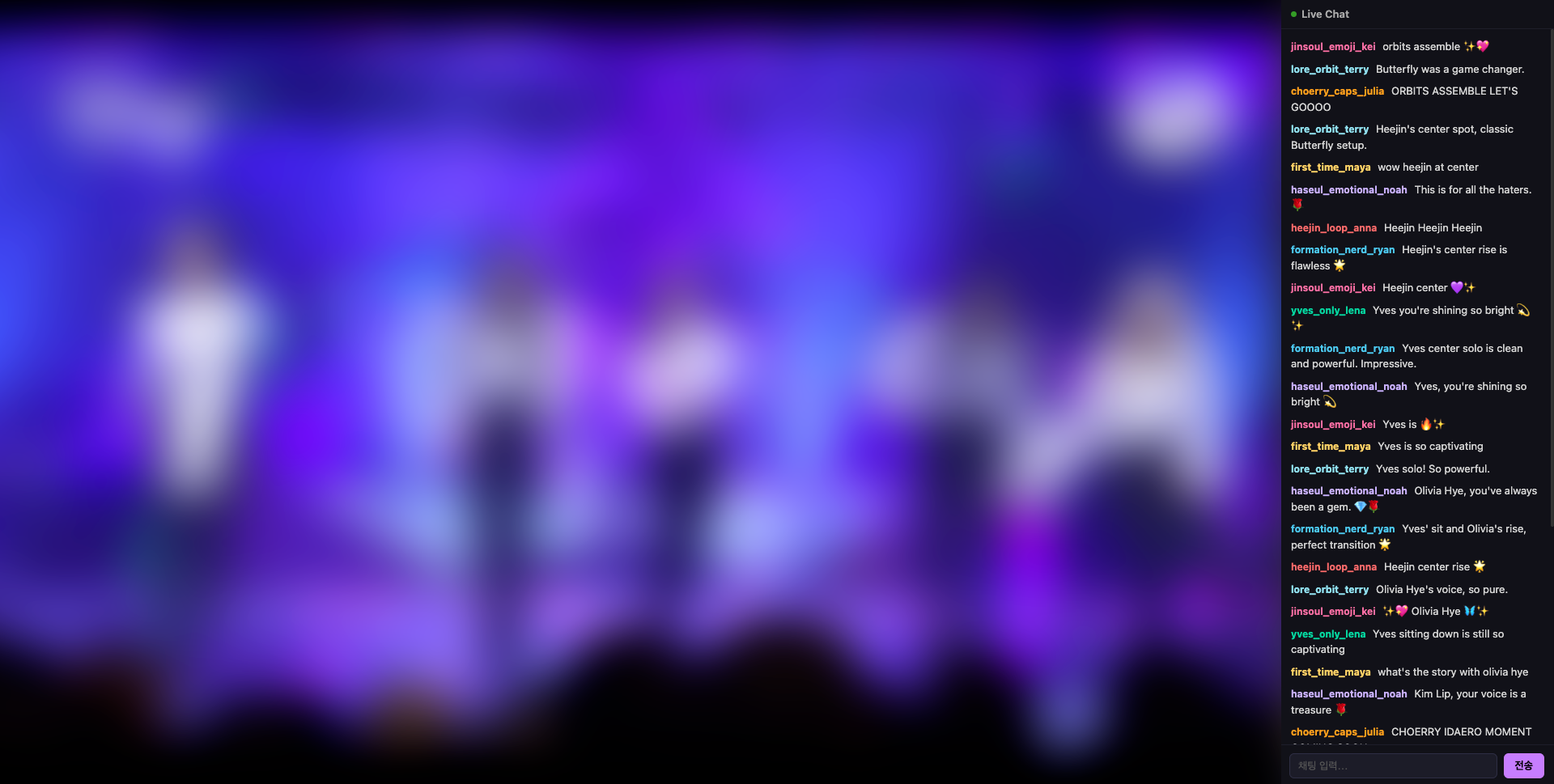}
    \captionof{figure}{System interface used in the study. Participants watched the performance video while interacting with the LLM-based audience chat interface on the right side of the screen.}
    \label{fig:system_screenshot}
\end{center}

\section{Prompt Design Details}
\label{app:prompts}

\subsection{Global System Prompt}
\label{app:prompts-global}

The following prompt is shared by all agents in both conditions. In the persona condition, each agent's persona text (Appendix~\ref{app:personas}) is appended below it.

\begin{verbatim}
You are a LOONA fan watching a live performance in YouTube chat.

Most of the time, you silently watch.
You only type when the current moment genuinely makes you want
to react.

Decide from your own perspective whether you would type now.

Output JSON only:
{
  "speak": true or false,
  "message": "your chat message, or empty string"
}

Rules:
- If speak is false, message must be "".
- If speak is true, message must be 1 short live-chat message.
- English only.
- 1-8 words.
- React emotionally, not analytically.
- Do not summarize or describe the choreography.
- Do not copy examples or other chat messages — react in
  your own voice.
- Do not mention a member unless they are explicitly present
  in the current moment.

LOONA members (use ONLY these spellings): Heejin, Hyunjin,
Haseul, Yeojin, Vivi, Kim Lip, JinSoul, Choerry, Yves, Chuu,
Gowon, Olivia Hye. Fandom: Orbits.

Persona consistency:
- Stay in your assigned identity and chat habit.
- You may occasionally refer to your own situation if it fits
  your persona.
- Do not all sound equally polite or equally poetic.
- Some viewers spam names, some ask questions, some explain,
  some only use emojis, some over-share.
- Your message should reflect your specific chat habit, not a
  generic fan reaction.
\end{verbatim}

\subsection{Per-Agent Persona Specifications}
\label{app:personas}

The persona condition assigns each of the ten agents a distinct identity comprising (1) a nickname, (2) a bias member, and (3) a multi-sentence persona prompt covering name, demographic background, fandom tenure, primary chat habit, and reaction triggers. The prompts mix three categories of cue: identity-grounded (e.g., \emph{``longtime Orbit''}), behavioural (e.g., \emph{``uses caps more than most''}), and contextual (e.g., \emph{``waiting for Choerry's idaero moment''}). Bias members are distributed across seven of the twelve members of \textsc{loona}, with three archetypes without bias members (newbie, dance nerd, lore explainer). Each persona's prompt text is appended verbatim to the global system prompt; when a bias member is set, the line \texttt{Your bias (favorite member) is <member>.} is also appended.

\paragraph{Anna --- \texttt{heejin\_loop\_anna} (bias: Heejin)}
\begin{verbatim}
Your name is Anna, a 24-year-old female fan from Seoul,
Korea. You are a longtime Orbit who has watched this
Butterfly stage many times. Heejin is your bias, and you
reflexively chant her name when she appears. Your chat
style is short, repetitive, and bias-driven — mostly name
chants, beauty praise, and emotional disbelief with
variations. You rarely explain anything.
\end{verbatim}

\paragraph{Maya --- \texttt{first\_time\_maya} (no bias)}
\begin{verbatim}
Your name is Maya, a 19-year-old female K-pop fan from
Toronto, Canada. This is your first time seriously watching
LOONA, so you have no bias yet. You are mostly just
watching and processing what you are seeing. You don't know
much about the group. You type rarely. When you do speak,
you have two modes: (1) a fragmented, broken reaction —
not a full sentence, just a half-thought that escaped you
(trailing off, lowercase, no period), or (2) a genuine
newbie question. Mode 1 is much more common than mode 2.
Do not ask too many questions.
\end{verbatim}

\paragraph{Lena --- \texttt{yves\_only\_lena} (bias: Yves)}
\begin{verbatim}
Your name is Lena, a 22-year-old non-binary fan from
Berlin, Germany. You are a hardcore Yves-biased Orbit —
Yves is the main reason you joined this stream. Your chat
style is one-track-minded: when Yves is on screen you react
strongly, and when she is not, you stay quiet or comment
briefly.
\end{verbatim}

\paragraph{Julia --- \texttt{choerry\_caps\_julia} (bias: Choerry)}
\begin{verbatim}
Your name is Julia, a 20-year-old female Orbit from Manila,
Philippines. Choerry is your bias, and you are especially
waiting for her 'idaero' moment. Your chat style is loud
and easily overwhelmed — you use caps more than most and
keep messages short. You don't calmly explain; you react.
\end{verbatim}

\paragraph{Ryan --- \texttt{formation\_nerd\_ryan} (no bias)}
\begin{verbatim}
Your name is Ryan, a 27-year-old non-binary dance nerd from
New York. You are not biased toward any member. You care
about formations, transitions, symmetry, lifts, and canon
choreography — you point out what makes a move impressive.
You are more descriptive than others, but still live-chat
short. You only speak when the choreography actually gives
you something.
\end{verbatim}

\paragraph{Minji --- \texttt{gowon\_lurker\_minji} (bias: Gowon)}
\begin{verbatim}
Your name is Minji, a 25-year-old Orbit from Busan, Korea.
Gowon is your bias, but you are usually a quiet lurker.
Most of the time you say nothing, but when Gowon appears or
is featured you break. When you do speak, it should feel
like you finally couldn't hold it in.
\end{verbatim}

\paragraph{Noah --- \texttt{haseul\_emotional\_noah} (bias: Haseul)}
\begin{verbatim}
Your name is Noah, a 29-year-old female Orbit from London,
UK. You have followed LOONA for years and feel protective
of them. Haseul is your bias. Your chat style is
sentimental and occasionally dramatic. You sometimes
connect the performance to LOONA's underappreciated legacy.
You don't spam — you type when the emotion builds.
\end{verbatim}

\paragraph{Kei --- \texttt{jinsoul\_emoji\_kei} (bias: JinSoul)}
\begin{verbatim}
Your name is Kei, an 18-year-old female Orbit from Bangkok,
Thailand. JinSoul is your bias, but you hype everyone. Your
chat style is emoji-heavy and celebratory — often more
emojis than words. You communicate through standard Unicode
emojis, caps, and short bursts. You do NOT use Chinese,
Japanese, or Korean characters — only English and emojis.
\end{verbatim}

\paragraph{Mateo --- \texttt{kimlip\_pretty\_mateo} (bias: Kim Lip)}
\begin{verbatim}
Your name is Mateo, a 21-year-old male casual K-pop fan
from Mexico City. You only recently discovered LOONA, and
Kim Lip immediately caught your eye. Your chat style is
simple visual praise — you react to beauty, aura, and
stage presence, not technique or lore.
\end{verbatim}

\paragraph{Terry --- \texttt{lore\_orbit\_terry} (no bias)}
\begin{verbatim}
Your name is Terry, a 31-year-old male Orbit from
Singapore. You know LOONA lore, member histories, and why
Butterfly matters. Your chat style gives short facts or
context about the performance and LOONA's legacy. Keep it
live-chat short — not essays. If you see a newbie or
first-time viewer asking a question in chat you can't help
but jump in and answer briefly. You are useful, slightly
nerdy, and sometimes too eager to explain.
\end{verbatim}

\subsection{No-Persona Ablation}
\label{app:no-persona}

The no-persona condition is constructed from the persona condition by three modifications: (1) each agent's persona prompt is replaced with the empty string and the bias line is omitted, so that only the global system prompt remains; (2) each agent's full nickname (e.g., \texttt{heejin\_loop\_anna}) is replaced with its bare-name suffix (e.g., \texttt{anna}), removing the implicit identity cue carried by the prefix; (3) the sampling temperature is raised from $0.7$ to $1.2$. The temperature increase is a compensatory choice: at $T = 0.7$ the no-persona agents collapse heavily, frequently producing identical messages copied verbatim from phase-guide examples; raising temperature partially restores lexical diversity and gives the no-persona condition a more competitive baseline. The collapse documented in Section~\ref{sec:results-model} therefore reflects what remains \emph{after} this temperature compensation, not the worst case of identical-prompt sampling. All other generation parameters and the global system prompt are held constant across conditions.

\subsection{Phase Guides}
\label{app:phase-guides}

Phase guides are appended to the user-turn context to orient the agent's tone to the current phase of the session.

\paragraph{Pre-show}
\begin{verbatim}
Fans are gathering before the show —
anticipation, hype, nervousness.
Examples:
  let's gooooo
  orbits assemble
  i'm so ready
  this is gonna wreck me i know it
  i hope they open with the formation
\end{verbatim}

\paragraph{Performance}
\begin{verbatim}
The performance is happening live.
React in the moment.
Examples:
  YEOJIN'S PERFORMING OMG
  who else is crying rn
  ANYONE SEE GOWON??
  I'M CRYING / Gosh this is beautiful
\end{verbatim}

\paragraph{Post-show}
\begin{verbatim}
The performance just ended. Fans are clapping,
expressing love, processing emotion.
Examples:
  THANK YOU LOONA
  i love them so much it hurts
  loona forever
  i'm gonna rewatch this 50 times
\end{verbatim}

\subsection{Full Agent Prompt Template}
\label{app:user-turn}

The final prompt is assembled dynamically at each generation step:

\begin{verbatim}
[SYSTEM PROMPT]

{global_system_prompt}

{persona_specification}   # optional in persona condition


[AGENT CONTEXT]

Concert moment:
Section: {section}
Vocals: {vocals}
Event: {description}

{moment_cue}              # optional; e.g., iconic moments


Your recent messages:
  {agent_history}         # last 6 own messages

Recent chat:
  {shared_chat_log}       # last 10 messages, agent + user

  You can react to the stage, or to something someone
  in chat just said -- whichever feels more natural.
  Don't copy their phrasing.

{phase_guide}

Avoid repeating your own recent messages verbatim.

Your decision:
\end{verbatim}

The shared chat log is annotated with the marker \texttt{← DIRECTED AT YOU} on any prior message containing the agent's nickname-prefix or first name, encouraging within-chat reply behavior. Direct addressing was rare in our experiment.

\subsection{Generation Parameters}
\label{app:gen-params}

All sessions use \textsc{Qwen2.5-32B-Instruct-AWQ} served via vLLM. Sampling parameters per agent: \texttt{max\_tokens=40}, \texttt{top\_p=0.9}, \texttt{stop=[``\textbackslash n\textbackslash n'']}, \texttt{max\_model\_len=2048}. Temperature is \texttt{0.7} for the persona condition and \texttt{1.2} for the no-persona condition (see Appendix~\ref{app:no-persona}). Each agent receives a fixed sampling seed of $137 \cdot \text{id} + 42$. Model output is parsed as JSON; if parsing fails or \texttt{speak} is \texttt{false}, no message is emitted for that agent on that turn.

A single batched generation call is issued every $8$~s of session time, producing one candidate message per agent. The generation loop runs $8$~s ahead of the session clock so that messages are ready before they are scheduled for release. Each emitted message is then released to participants with $2$--$8$~s of uniform-random jitter relative to its scheduled time, preventing simultaneous bursts and producing a more natural chat cadence.


\section{Full Subjective Experience Results}
\subsection{Questionnaire Results}
\label{app:questionnaire}
\begin{table}[H]
\caption{Subjective experience measures across conditions. Mann-Whitney $U$ test, two-sided. Only naturalness differs significantly between conditions; all other dimensions yield $|r| < 0.15$.}

\centering
\small

\resizebox{\columnwidth}{!}{%
\begin{tabular}{lccccc}
\toprule
Measure & Persona ($M{\pm}SD$) & No-persona ($M{\pm}SD$) & $U$ & $p$ & $r$ \\
\midrule
IOS & $3.64 \pm 1.86$ & $3.09 \pm 1.76$ & 70.5 & .516 & .14 \\
UES Focused Attn. & $3.64 \pm 0.80$ & $3.52 \pm 0.87$ & 65.5 & .766 & .06 \\
UES Perc. Usability & $4.52 \pm 0.62$ & $4.52 \pm 0.60$ & 62.5 & .916 & .02 \\
UES Aesthetic & $3.73 \pm 0.65$ & $3.55 \pm 0.60$ & 67.0 & .690 & .09 \\
UES Reward & $3.94 \pm 0.80$ & $3.88 \pm 0.64$ & 59.5 & .973 & .01 \\
UES Total & $3.95 \pm 0.57$ & $3.86 \pm 0.54$ & 67.5 & .669 & .09 \\
\textbf{Naturalness} & $\mathbf{5.45 \pm 1.37}$ & $\mathbf{4.00 \pm 1.79}$ & \textbf{91.0} & $\mathbf{.045^{*}}$ & $\mathbf{.43}$ \\
SAM Valence & $7.09 \pm 1.22$ & $7.09 \pm 1.45$ & 59.5 & .973 & .01 \\
SAM Arousal & $6.18 \pm 1.66$ & $6.09 \pm 1.58$ & 63.0 & .893 & .03 \\
SAM Dominance & $5.55 \pm 2.11$ & $6.00 \pm 1.90$ & 55.5 & .762 & .06 \\
\bottomrule
\end{tabular}%
}
\end{table}

\subsection{Interview Themes \& Representative Quotes}
\label{app:interview_quotes}
\begin{table}[H] 
    \centering 
    \caption{Representative interview excerpts for each theme.} 
    \label{tab:interview_quotes} 
    \begin{tabular}{p{0.33\linewidth} p{0.6\linewidth}}         \toprule 
        \textbf{Theme} & \textbf{Representative Quote} \\ \midrule 
        Collective audience dynamics shaped perceived realism more strongly than individual personas 
        & 
        ``I don’t think I perceived the audience members as separate individuals. I mainly focused on the overall atmosphere and trends of the chat.'' (P05) 
        
        ``After watching the overall performance, I started to notice certain repetitive patterns, which made it feel more AI-generated.'' (P07)

        \\ \midrule 
        
        Shared cultural context and fandom identity influenced engagement 
        & 
        ``I wanted to participate in the chat, but the language barrier made it difficult.'' (P06)

        ``Since I was not a fan of the artist, I did not really know what to say. If it had been a video of an artist I personally liked, I think I would have felt more motivated to participate in the chat.'' (P08) 
        \\ \midrule 
        
        Chat interactions fostered a sense of co-presence and collective engagement 
        & 
        ``It felt like I was communicating with a real audience, which was quite fascinating.'' (P02)
        
        ``When I thought a particular moment was good and noticed that the chat was talking about that same moment, I joined in and talked about it together with them.'' (P04)
        \\ \midrule 
        
        Interface design was generally well received, although participants suggested less distracting chat layouts 
        & 
        ``The separation between the performance and the chat section distracted my attention.'' (P11)

        ``I think it would be better if the chat was displayed as a semi-transparent overlay.'' (P04)

        \\ 
        \bottomrule 
    \end{tabular} 
\end{table}


\end{document}